\documentclass[
journal=jctcce, %
manuscript=article, layout=onecolumn]{achemso}
\usepackage{verbatim}
\usepackage[version=3]{mhchem} %
\usepackage{color}
\usepackage{amsmath, bm}
\usepackage{graphicx}
\usepackage{physics}

\newcommand{\CX}{\ensuremath{\mathcal{X}}}

\author{Felix Musil}
\email{felix.musil@epfl.ch}
\author{Michael J.~Willatt}
\email{michael.willatt@epfl.ch}
\affiliation{Laboratory of Computational Science and Modeling, IMX, \'Ecole Polytechnique F\'ed\'erale de Lausanne, 1015 Lausanne, Switzerland}

\author{Mikhail A.~Langovoy}
\affiliation{Machine Learning \& Optimization Laboratory, IC, \'Ecole Polytechnique F\'ed\'erale de Lausanne, 1015 Lausanne, Switzerland}

\author{Michele Ceriotti}
\affiliation{Laboratory of Computational Science and Modeling, IMX, \'Ecole Polytechnique F\'ed\'erale de Lausanne, 1015 Lausanne, Switzerland}

\usepackage{bm}

\newcommand\by{\mathbf{y}}
\newcommand\bz{\mathbf{z}}

\newcommand\R{\textrm{R}}
\newcommand\GPR{\textrm{GPR}}

\newcommand{\angstrom}{\text{\normalfont\AA}}

\let\oldmaketitle\maketitle
\let\maketitle\relax

\title[Errors in ML]
{Fast and Accurate Uncertainty Estimation\\ in Chemical Machine Learning}

\begin{document}

\oldmaketitle
\begin{abstract}
We present a scheme to obtain an inexpensive and reliable estimate of the uncertainty associated with the predictions of a machine-learning model of atomic and molecular properties. The scheme is based on resampling, with multiple models being generated based on sub-sampling of the same training data.  The accuracy of the uncertainty prediction can be benchmarked by maximum likelihood estimation, which can also be used to correct for correlations between resampled models, and to improve the performance of the uncertainty estimation by a cross-validation procedure. In the case of sparse Gaussian Process Regression models, this resampled estimator can be evaluated at negligible cost.  We demonstrate the reliability of these estimates for the prediction of molecular energetics, and for the estimation of nuclear chemical shieldings in molecular crystals. Extension to estimate the uncertainty in energy differences, forces, or other correlated predictions is straightforward.  This method can be easily applied to other machine learning schemes, and will be beneficial to make data-driven predictions more reliable, and to facilitate training-set optimization and active-learning strategies.
\vspace{0.5cm}
\end{abstract}

{\let\thefootnote\relax\footnote{$^*$These authors contributed equally to this manuscript}}

\section{Introduction}

Much research over the last  decade has explored the applicability of machine learning methods to materials science and computational chemistry. In particular, a very promising line of research involves using statistical regression models (artificial neural networks, kernel ridge regression, decision trees, etc.) to predict quantum mechanical properties of atomistic systems (energy, response functions, density, etc.) based on a small number of reference calculations, using only the atomic structure as input\cite{Behler2007, Bartok2010, Rupp2011,Ward2016}.
The main reason these methods have gained so much traction in recent years is their promise to achieve high accuracy predictions at an affordable computational cost.\cite{VonLilienfeld2018} For example, it is possible to train a neural network to predict the formation energy of a crystal by providing it with a training set of atomic structure-energy pairs, derived from an electronic structure calculation. While the accuracy of the neural network is ultimately throttled by the accuracy of the underlying electronic structure method, it can be driven towards this theoretical maximum by increasing the size of the training set.\cite{Amari1993,Muller1996} 
Then, the trained neural network competes with the accuracy of the underlying electronic structure method despite having a computational expense that might be many orders of magnitude smaller. This enables a huge array of investigations that would otherwise be prohibitively expensive.\cite{Behler2016, Smith2017, Bartok2018, Deringer2017, Bartok2017, Kobayashi2017}

One of the possible approaches to quantify the accuracy of a machine learning model when presented with new inputs (the generalization error), involves measuring the residual error on a set of input-observation pairs (the test set) that are deliberately excluded from the training phase. 
These residual errors might be combined into a single score for the model, e.g. the Root Mean Square Error (RMSE) or Mean Absolute Error (MAE), which provides an estimate of the magnitude of the expected residual for an arbitrary input on average. 
Scores like the RMSE and MAE are undoubtedly useful guides, but one would often like an estimate of the error or uncertainty associated with a particular input.\cite{fox2015applied, Tibshirani1996, Peterson2017,Bartok2018, Bishop1997, Goldberg1997, Nix1994}
Roughly speaking, one would like to know when the model is interpolating and when it is extrapolating (and thus likely to be less reliable). Not only can such a measure of prediction uncertainty allow the computational chemist to draw conclusions more confidently from the model, it can also direct the construction of the training set by highlighting important regions of the input space that are underrepresented.\cite{Rasmussen2006, Titsias2009} 

In this article we focus on Gaussian Process Regression (GPR)\cite{Rasmussen2006} for the prediction of atomic structure properties with the well-established Smooth Overlap of Atomic Positions\cite{Bartok2013a, Deringer2017, De2016, Bartok2017} (SOAP) framework to generate the kernels. 
In the GPR framework a prediction is the mean of a normal distribution that depends on the training set, and the standard deviation of this distribution provides an uncertainty estimate for each prediction. 
We compare this uncertainty estimator with another that is based on sub-sampling\cite{Anderson1986, Politis1999} of the training set, where multiple models are trained on different portions of the training set and the distribution of predictions across the models is used to estimate the prediction uncertainty. 
We discuss ways to assess the relative performance of different uncertainty estimators, and to improve the performance of an estimator by a calibration procedure based on cross-validation. We demonstrate this framework by assessing the accuracy of SOAP-GPR predictions of formation energies in the QM9 data set\cite{Ramakrishnan2014} and \textsuperscript{1}H NMR chemical shieldings in the CSD data set.\cite{Groom2016,Paruzzo2018}

\section{Methods}

We start by giving a short summary of the GPR framework, the projected process (PP) approximation, and the associated uncertainty estimators. These are well-known results, which can also be found e.g. in Refs.~\citenum{Rasmussen2006, Quinonero-candela2005}, but we report them here for completeness and to introduce our notation. 
We then discuss resampling estimators of sample statistics, and the use of maximum likelihood to compare and optimize the fitness of different statistical models.

\subsection{Gaussian Process Regression}

In regression, a target property $y$ of an input $\mathcal{X}$ is often assumed to be contaminated with noise,
\begin{equation}
    y(\mathcal{X}) = f(\mathcal{X})+\epsilon.
\end{equation}
In the GPR framework, we further assume that $\epsilon\sim\mathcal{N}(0,\sigma)$ is Gaussian random noise and the underlying function we wish to model $f\sim\mathcal{GP}(0,K)$ is a Gaussian Process.

Effectively, a GPR model predicts the target property $y$ of an input $\mathcal{X}$, given a training set containing $N$ input and observation pairs $D \equiv \{(\mathcal{X}_{i}, y_{i})\}$, as a linear combination of basis functions,
\begin{equation}
    y(\mathcal{X}) = \sum_{i} x_{i} k(\mathcal{X}_{i}, \mathcal{X}),
\end{equation}
where the inputs $\{\mathcal{X}_{i}\}$ are those for which the target property has been observed. The function that serves as the basis $k(\mathcal{X}, \mathcal{X}')$ is called the kernel and should represent the similarity between its inputs, so that it is large if two inputs are similar and small if they are not. 

Exact GPR models become computationally prohibitive for large training sets, so many sparse approximations have been developed to overcome this limitation in recent years.\cite{Rasmussen2006,Liu2018} 
We use the Projected Process (PP) approximation, also called Projected Latent Variables, throughout this paper.\cite{Seeger2003,Rasmussen2006} 
A subset $M$ of the $N$ training inputs is selected as an `active' or `representative' set, and it is used as the new basis for the regression, enhanced by its correlations with the full training set.
The utility of compressing the model in this way is strongly dependent on the choice of the representative set. Possible approaches include Farthest Point Sampling (FPS),\cite{Ceriotti2013b,Imbalzano2018} to maximize the dissimilarity of the inputs in the representative set, and the use of CUR to minimize the effect of the PP approximation on the kernel matrix of the entire training set.\cite{bart-csan15ijqc}
The approximate GPR prediction for a new input $\mathcal{X}$ and the variance associated with this prediction are:
\begin{equation}
    \begin{split}
    y_{PP}(\mathcal{X}) &= K_{\mathcal{X}M} \Tilde{K}^{-1} K_{MN} \bm{y},\\
    \sigma_{PP}^2(\mathcal{X}) &=  \sigma^{2} + K_{\mathcal{X}\mathcal{X}} - Q_{\mathcal{X}\mathcal{X}} + K_{\mathcal{X}M} \Tilde{K}^{-1} K_{\mathcal{X}M}^T,
    \end{split}
    \label{eq:gpr-pp}
\end{equation}
where $K_{NM}$ and $K_{\mathcal{X}M}$ are the kernel matrices between the $M$ active inputs and the $N$ training inputs and the new input $\mathcal{X}$ respectively, $Q_{\mathcal{X}\mathcal{X}} = K_{\mathcal{X}M} K_{MM}^{-1} K_{\mathcal{X}M}^{T}$ and $\Tilde{K}=\left( K_{MM}+\sigma^{-2}K_{NM}^T K_{NM} \right)$.

The probabilistic nature of the GPR model provides an uncertainty estimate of its predictions: if the Gaussian process assumption is correct, then the interval $[y(\mathcal{X}) - \sigma_{PP}(\mathcal{X}), y(\mathcal{X}) + \sigma_{PP}(\mathcal{X})]$ will contain the target 67\% of the time. If the interval is particularly large, then the prediction is untrustworthy (though in many cases the actual error might be small), and the training set should be extended with points close to $\mathcal{X}$.

\subsection{Resampling}

Another approach to estimate the uncertainty associated with a prediction involves creating a family of models based on the same input data, which are representative of the statistical error associated with the finite amount of available training inputs.\cite{Efron1979, Tibshirani1996, Anderson1986, Politis1999} Here we will discuss bootstrapping and sub-sampling techniques, which are applicable to any predictive statistical model.

Given the original training set $D$ of size $N$, one creates $N_{\R}$ new data sets by drawing $n$ input-observation pairs from $D$. In bootstrapping, the input-observation pairs are drawn from $D$ with replacement and $n = N$, whereas in sub-sampling the selection is performed without replacement and $n < N$. Models trained independently on this ensemble of resampling data sets produce a fully non-parametric estimate of the distribution of the prediction for an input $\CX$, $P\left(y\middle|\mathcal{X}\right)$, whose moments can be calculated, e.g.
\begin{equation}
\begin{split}
    y_{RS}(\mathcal{X}) =& \, \frac{1}{N_{\R}}\sum_{i} y^{(i)}(\mathcal{X}) \\
    \sigma^{2}_{RS}(\mathcal{X}) =& \frac{1}{N_{\R}-1}\sum_{i} \left[y^{(i)}(\mathcal{X}) - y_{RS}(\mathcal{X}) \right]^{2},
\end{split}
\label{eq:resamp-mu-sigma}
\end{equation}
where $y^{(i)}(\mathcal{X})$ is the prediction for the $i$\textsuperscript{th} resampling model. An advantage of this family of methods is that the ensemble of predictions $\left\{y^{(i)}(\mathcal{X})\right\}$ provides a full characterization of the error statistics which makes it possible to evaluate a non-parametric empirical model of $P\left(y\middle|\mathcal{X}\right)$. What is more, when considering multiple inputs $\left\{X_n\right\}$, the ensemble of predictions relates to the fully correlated prediction distribution $P\left(\bm{y}\middle|\left\{\mathcal{X}_n\right\}\right)$, making it trivial to estimate the uncertainty for any combination of the predictions. 

In bootstrapping, the procedure is called the pairs resampling algorithm (as opposed to the bootstrap residuals resampling algorithm),\cite{Tibshirani1996, fox2015applied} and it is commonly used in machine learning to construct committee models\cite{Breiman1996} and estimate prediction uncertainties. 
In the context of uncertainty estimation, the bootstrap variance $\sigma^{2}_{RS}(\mathcal{X})$ is sometimes used to estimate uncertainties in predictions from neural networks, where it has been found to be more reliable than alternative estimators because the variability of the predictions with respect to the initialization parameters of the neural network is incorporated automatically.\cite{Tibshirani1996, Heskes1997}

Bootstrapping generates random samples of the right size $N$ from the wrong distribution. On the other hand, sub-sampling generates random samples of the wrong size $n < N$ from the right distribution, provided $n \ll N$.\cite{Politis1999} There are a variety of approaches to correct for this shortcoming of the sub-sampling approach. The most common is to assume a power law relationship between the statistic one is interested in and the size of the sub-sample $n$. Linear regression for a variety of sub-sample sizes then allows one to infer the exponent in the power law and extrapolate to the $n \rightarrow N$ limit.\cite{Politis1999}

Instead, to extrapolate to the $n \rightarrow N$ limit, we apply a linear (${\sigma}\rightarrow\alpha\sigma$) or non-linear (${\sigma}\rightarrow\alpha\sigma^\gamma$) transformation to the predicted uncertainties. For resampled estimators this can be implemented as a transformation of the ensemble of predictions (see Eq.~\ref{eq:rs-nonlinear-scaling}), which allows us to correct the distribution of predictions without assuming a particular functional form. 
The parameters of this transformation can be determined using maximum likelihood estimation described below. 
This procedure can be applied to sub-sampling and bootstrapping alike, but the criteria we use to assess uncertainty estimates suggest that bootstrapping offers no advantage over sub-sampling for uncertainty estimation with the GPR PP models presented here. %
Since sub-sampling is simpler and computationally cheaper, we focus on sub-sampling in the remainder of this article.

A practical concern regarding resampling algorithms is that they require $N_{\R}$ models to be trained, which increases the computational cost $N_{\R}$-fold.\cite{Tibshirani1996, Peterson2017} However, this added computational cost is associated with the training phase, whereas calculating the GPR variance is expensive in the testing phase. In most situations where one would like an uncertainty estimate for each prediction, an extended training phase is often preferable over an increased cost of making predictions. 
Moreover, if one exploits the model compression scheme (GPR PP) outlined earlier, where the training set is partitioned into active and passive components, then the computational expense of training and testing can be reduced significantly for both the resampling and GPR approaches to uncertainty estimation.
Furthermore, if one uses the same representative set for all models then the cost of predicting the uncertainty for a resampling estimator becomes effectively zero. In fact, one only needs to compute once the kernel between the new input and the representative set (which is typically the expensive step), and evaluating multiple models requires only the calculation of $N_\R$ scalar products. This is in stark contrast to similar approaches based on neural networks~\cite{Peterson2017,Behler2014} in which the evaluation of multiple models entails a substantial overhead. 
The PP approximation is, however, known to be detrimental to the quality of the GPR variance and, to a lesser extent, the prediction. 

\subsection{Log-likelihood assessment of uncertainty estimates}

Assessing the prediction accuracy of a machine learning model is straightforward, as it suffices to compute some average of the prediction errors $y_{n} -y(\mathcal{X}_{n})$ for an appropriate validation/test set of points.  
However, how should one assess the quality of a model that provides an estimate of the uncertainty $\sigma(\mathcal{X})$ as well as of the value of the property $y(\mathcal{X})$? 
Here, we use for this purpose a log-likelihood estimator that has also been adopted for the same purpose in some classical works on statistical regression~\cite{Bishop1997,Ibragimov_1978}.

In a nutshell, we assume that the true values of the properties $\bm{y}$ associated with the test structures $\left\{\mathcal{X}_n\right\}_{n=0,1,...}$ are uncorrelated and follow a Gaussian probability distribution,
\begin{equation}
P\left(\bm{y}\middle|\left\{\mathcal{X}_n\right\}_{n=0,1,...}\right)=\prod_{n=0}^N \frac{1}{\sqrt{2\pi\sigma^2(\mathcal{X}_n)}} \exp \left(-\frac{(y_n - y(\mathcal{X}_n))^2}{2{\sigma}^2(\mathcal{X}_n)} \right),
\end{equation}
whose means $y(\mathcal{X}_n)$ and variances $\sigma^2(\CX_n)$ are determined with a statistical model -- a Gaussian Process or a committee of Gaussian Processes in the present work (see Eqs.~(\ref{eq:gpr-pp}) and (\ref{eq:resamp-mu-sigma}) respectively). 
The match between the predictions and the actual values of $y$ can be quantified by summing the logarithms of $P\left(y\middle|\mathcal{X}\right)$ over an appropriate test set -- corresponding to the logarithm of the probability that the true targets are a realization of the model,
\begin{equation}
LL = \frac{1}{N_\text{test}}\sum_{\mathcal{X}_n\in\text{test}} \log P\left(y_{n}\middle|\mathcal{X}_n\right).
\label{eq:ll}
\end{equation}
When using the same test set to compare two models that only differ by the uncertainty estimate, the best model will yield the highest value of $LL$. Note that a more general discussion of the likelihood for Gaussian probability distributions can be found in Ref.~\citenum{Bishop2006}.

\subsection{Maximum likelihood estimation for scaling uncertainty estimates}

Since sub-sampling models are trained with $n < N$ input-observation pairs, the distribution of predictions about the reference over a set of sub-sampling models is likely to be too broad or narrow in general. 
If we assume that the distributions for different inputs are broadened or narrowed by roughly the same amount, then this distortion can be corrected by scaling each distribution (or its moments) by the same constant. 
The same approach can of course be applied to the GPR variances, to correct for the detrimental effect of a small representative set, and also to bootstrapping to correct for artificial correlations between the resampled models.

Maximum likelihood estimation is a straightforward means of determining the constant. 
We suppose the reference prediction $y(\mathcal{X}_{n})$ that corresponds to $\mathcal{X}_{n}$ is normally distributed about the target $y_{n}$ with a variance $\sigma^{2}(\mathcal{X}_{n})$, which might correspond to GPR or resampling. 
Since this supposes the targets are uncorrelated, the total log likelihood of the targets is a sum of log likelihoods, 
\begin{equation}
    LL(\bz) = -\frac{1}{2} \sum_{n} \frac{z_{n}^{2}}{\sigma^{2}(\mathcal{X}_{n})}  
    + \log (2\pi \sigma^{2}(\mathcal{X}_{n})),
\end{equation}
where $z_{n} = y_{n} - y(\mathcal{X}_{n})$. In order to tune all the variances, we take $\sigma^{2}(\mathcal{X}_{n}) \to v \sigma^{2}(\mathcal{X}_{n})$ so the log likelihood becomes
\begin{equation}
    LL(\bz, v) = -\frac{1}{2} \sum_{n} \frac{z_{n}^{2}}{v \sigma^{2}(\mathcal{X}_{n})} 
    + \log (2\pi v \sigma^{2}(\mathcal{X}_{n})),
\end{equation}
The parameter $v$ can be optimized on a validation set of size $N_{\textrm{val}}$, where the residuals $\bz$ are known. Demanding that the log likelihood is maximized over the validation set leads to
\begin{equation}
    v_{0} = \frac{1}{N_{\textrm{val}}} \sum_{n} \frac{z_{n}^{2}}{\sigma^{2}(\mathcal{X}_{n})}
\end{equation}
We then use the parameter $v_{0}$ to scale the GPR variances or resampling distributions when the model is later used for testing.
Note that (as we will discuss in the results section) this scaling of the predicted variance is just one of the possible strategies that can be used to increase the accuracy of the uncertainty estimation. For instance, one can apply a more general transformation of $\sigma^2(\mathcal{X})$, computed from the GPR estimator or resampling. 

As is often the case, one should be wary of overfitting. While we mitigate this risk by a cross-validation procedure, in the context of maximum likelihood estimation it is customary to introduce a penalty for the complexity of the model. Commonly used techniques, such as the Bayesian Information Criterion \cite{Schwarz_1978_BIC}, or the Akaike Information Criterion \cite{Akaike_1974_AIC}, also allow for an information-theoretic interpretation that justifies a comparison between models of different complexity.
For the Gaussian process case, the scaling factor $v_{0}$ is the scalar product norm of the score vector for the sub-problem with $N_\text{val}$ observations. The covariance matrix is the Gram matrix of the score vector. Therefore, in view of the Cauchy-Schwarz inequality for scalar products in Hilbert spaces, $v_{0}$ serves as a normalizing factor for the $LL$. This implies that $LL(\bz, v_0)$ can be also used as a likelihood-based test statistic for testing the above assumption of uncorrelated normally distributed reference predictions \cite{Langovoy_2007}. 

In a demanding, real application, removing input-observation pairs from the training set might be overly wasteful. In Ref.~\citenum{Heskes1997}, Heskes points out that in a randomly-resampled data set $D_{i}$, many of the input-observation pairs in $D$ will be absent and can thus be used for validation. An attractive way of determining $v_{0}$ without explicitly constructing a validation set is therefore the following,
\begin{equation}
    y_\text{INT}(\mathcal{X}_n) = \frac{1}{N_{\R}(\mathcal{X})} \sum_{\substack{i \\ \mathcal{X} \not\in D_{i}}} y^{(i)}(\mathcal{X})
\end{equation}
\begin{equation}
    \sigma^{2}_\text{INT}(\mathcal{X}_n) = \frac{1}{N_{\R}(\mathcal{X}) - 1} \sum_{\substack{i \\ \mathcal{X} \not\in D_{i}}} \left[ y^{(i)}(\mathcal{X}) - y_\text{INT}(\mathcal{X}) \right]^{2},
\end{equation}
where $N_{\R}(\mathcal{X})$ is the number of resampling models that do not contain $\mathcal{X}$ in the training set. Then,
\begin{equation}
    v_{0} = \frac{1}{N_{\textrm{int}}} \sum_{n} \frac{z_{n}^{2}}{\sigma^{2}(\mathcal{X}_{n})}
\end{equation}
where the sum only includes an input $\mathcal{X}_{n}$ if it is absent from at least a few of the resampling data sets, $z_{n}$ is the difference between the target and the leave-one-out\cite{Cawley2006} prediction and $N_{\textrm{int}}$ is the number of absent inputs. It is straightforward to show that, as the size $N$ of the training set grows, the fraction of absent inputs for a random bootstrap sample tends to $e^{-1}$, while for sub-sampling it is always $1 - n/N$. This means for example that if one takes $N_{\textrm{R}} = 10$ and $n = N/2$, slightly more than 50\% of the training inputs are expected to be absent from at least five of the resampling models, and the size of the effective validation set for the procedure described above is therefore roughly half of the full training set. Note that this procedure cannot be applied to scaling GPR variances.

\subsection{Smooth Overlap of Atomic Positions}

We use the SOAP kernel in this work to compare atomic environments and structures, which is invariant under rotations, translations and permutations of identical atoms.\cite{Bartok2013, Bartok2013a, Bartok2017, De2016, Grisafi2018, Glielmo2018} For two atomic environments $\mathcal{X}_{i}$ and $\mathcal{X}_{j}$, the SOAP environmental kernel is given by an inner product between power spectra,
\begin{equation}
    k(\mathcal{X}_{i}, \mathcal{X}_{j}) 
    = \sum_{nn'l\alpha\alpha'} \bra{\mathcal{X}_{i}^{(2)}}\ket{\alpha n \alpha' n' l}\bra{\alpha n \alpha' n'}\ket{\mathcal{X}_{j}^{(2)}},
\end{equation}
where
\begin{equation}
\bra{\alpha n \alpha' n'}\ket{\mathcal{X}_{j}^{(2)}}
= \sqrt{\frac{8\pi^{2}}{2l+1}} \sum_{m} \bra{\mathcal{X}_{j}}\ket{\alpha n l m} \bra{\alpha' n' l m}\ket{\mathcal{X}_{j}},
\end{equation}
and the $\bra{\alpha n l m}\ket{\mathcal{X}_{j}}$ are linear expansion coefficients of the density associated with atoms of type $\alpha$ in the environment,
\begin{equation}
     \bra{\alpha n l m}\ket{\mathcal{X}_{j}} = \int d\mathbf{r} \, r^{2} \, \psi^{\alpha}_{\mathcal{X}_{j}}(\mathbf{r}) Y_{l}^{m}(\mathbf{r}) R_{n}(r).
\end{equation}
The $Y_{l}^{m}(\mathbf{r})$ are spherical harmonics, $R_{n}(r)$ are radial basis functions, and the density of atoms of type $\alpha$ in the environment is a linear combination of atom-centred Gaussians $g(\mathbf{r})$,
\begin{equation}
    \psi^{\alpha}_{\mathcal{X}_{j}}(\mathbf{r}; \mathcal{X}_{i}) = \sum_{j \in \alpha} g(\mathbf{r} - \mathbf{r}_{ij}) f_{\textrm{cut}}(r_{ij}).
\end{equation}
The cutoff function $f_{\textrm{cut}}(r_{ij})$ restricts the sum to atoms that are close to the atom at the centre of $\mathcal{X}_{i}$, for which $r_{ij} \equiv |\mathbf{r}_{i} - \mathbf{r}_{j}|$ is small. For the \textsuperscript{1}H NMR chemical shielding predictions and uncertainty analysis, the environmental kernel is used directly because chemical shieldings are associated with individual atoms. 
For the formation energy predictions and uncertainty analysis, the environmental kernel is averaged over all environments in each molecule, because the formation energy is a property of the entire molecule~\cite{De2016}. See Ref.~\citenum{Bartok2017} for a detailed discussion of SOAP for regression.

Note that, by default, SOAP kernels are dimensionless. For the GPR interpretation of the kernel as a covariance to make sense, they must assume the squared units of the property one wants to predict. This is easily achieved by taking
\begin{equation}
    K \frac{ \textrm{Var}[\by]} {\textrm{Tr}\left[K\right]} \to K,
\end{equation}
and the regularization parameter $\sigma^{2}$ must then be scaled by the same amount. This procedure has absolutely no effect on the prediction $y(\mathcal{X})$ but is essential to make the uncertainty estimate $\sigma^{2}_{\GPR}(\mathcal{X})$ dimensionally correct and therefore meaningful.

The SOAP framework has been adopted in many different applications in chemistry and materials science, including ligand binding classification, tensorial property prediction, property prediction for silicon, etc. Recently, Bartok \textit{et al.} presented a comprehensive analysis of prediction errors in a silicon potential which uses the SOAP environmental kernel as a basis.\cite{Bartok2018} They found that the GPR variance (subset of data approximation instead of PP approximation) provides a fair measure of the uncertainties associated with the potential for different silicon atom configurations, and they recommend it as a guide when expanding Gaussian approximation potential databases.

\subsection{Benchmark Datasets}

\subsubsection{Molecular Crystals dataset}

The CSD-61k data set\cite{Paruzzo2018} is based on 61k organic crystals extracted from the Cambridge Structural Database (CSD),\cite{Groom2016} containing only H, C, N and O atoms, and having fewer than 200 atoms in the unit cell. 
Out of the CSD-61k set, 500 randomly picked structures, the CSD-500 dataset, were optimized at the DFT+VdW level of theory.\cite{Paruzzo2018} 
To generate a benchmark database for this work, that includes a more diverse set of off-equilibrium environments, the structures from the CSD-500 dataset were randomly perturbed away from their optimal geometries, so that the Root Mean Squared Deviation (RMSD) between the positions of an optimized structure and its rattled counterparts is $0.25\angstrom$ and $0.5\angstrom$. We call this dataset, that contains 890 structures, the CSD-890-R. 
The \textsuperscript{1}H chemical shieldings were calculated using the Quantum Espresso package\cite{Varini2013,Giannozzi2009,Giannozzi2017}. 
We used PBE\cite{Perdew1996} ultrasoft pseudopotentials with GIPAW\cite{Pickard2001,Yates2007} reconstruction, H.pbe-kjpaw\_psl.0.1.UPF, C.pbe-n-kjpaw\_psl.0.1.UPF, N.pbe-n-kjpaw\_psl.0.1.UPF and O.pbe-n-kjpaw\_psl.0.1.UPF from the PSlibrary 0.3.1\cite{Kucukbenli2014}. 
The wave-function and charge density energy cut-offs were set to $100$ Ry and $400$ Ry respectively, the convergence threshold of the self consistent cycle is set to $10^{-12}$ Ry and a Monkhorst-Pack grid of k-points\cite{Monkhorst1976} corresponding to a maximum spacing of $0.06\angstrom$ in the reciprocal space.
The scalar chemical shieldings are obtained from the average of the diagonal of the chemical shielding tensor using a linear response wave-vector of $0.02$ bohrradius$^{-1}$ and a convergence threshold of $10^{-14}$ Ry$^2$ for the Green's function solver.

We randomly partitioned 30k of the H environments into 20k, 5k and 5k sets. One of the 5k sets was used for validation and the other was used for testing. We sorted the 20k training environments using FPS and used the 10k most diverse environments as the representative set for the PP approach. For sub-sampling we selected 16 random sub-samples of the training set for each sub-sample size. When using the training set in the log likelihood procedure described earlier, we only counted environments if they were absent from at least five realizations of each sub-sampling model.

\subsubsection{QM9 data set}

The QM9 data set\cite{Ramakrishnan2014} contains about 134k structures of small organic molecules optimized at the hybrid-DFT level\cite{Stephens1994} containing up to nine heavy atoms (C, N, O and F), in addition to hydrogen atoms. 
Following Ref.~\citenum{Ramakrishnan2014}, we removed all the 3,054 structures for which the SMILES string describing the molecule after relaxation differed from the starting configuration. We used as a target for the model the formation energies that include a zero-point energy correction.
We randomly partitioned 30k of the QM9 structures into 20k, 5k and 5k sets. One of the 5k sets was used for validation and the other was used for testing. We sorted the 361k environments present in the 20k training structures with FPS and used the 5k most distant environments as the representative set in the PP approach. For sub-sampling we selected 16 random sub-samples of the training set for each sub-sample size.

\section{Results and Discussion}

\subsection{Prediction of CSD \textsuperscript{1}H NMR chemical shieldings}

\begin{figure}[H]
    \centering
    \begin{tabular}{l r r r r r } 
    \hline\hline
    Size &  \multicolumn{1}{c}{LL (Raw)}   & \multicolumn{1}{c}{LL (CV)} & \multicolumn{1}{c}{LL (N-L)} & \multicolumn{1}{c}{LL (Int.)} & \multicolumn{1}{c}{LL (Int., N-L)}\\
    \hline
    1k  &    -0.554 & -0.398 & -0.356 & -0.398 & -0.356 \\
    5k  &    -0.389 & -0.381 & -0.338 & -0.382 & -0.337 \\
    10k &    -0.952 & -0.392 & -0.324 & -0.407 & -0.324 \\
    15k &    -4.102 & -0.394 & -0.312 & -0.470 & -0.320 \\
    18k &    -18.75 & -0.412 & -0.308 & -0.517 & -0.351 \\
    GPR PP & -8.771  & -0.255 & -0.251 & N/A    & N/A    \\  
    \hline
    \end{tabular}
    \captionof{table}{Log likelihood (LL) of predictions on the test set for different sub-sample sizes. After scaling the variances through maximum likelihood estimation -- internally (Int.) or on the validation set (CV) -- the final log likelihood is insensitive to the sub-sample size. A non-linear scaling of the uncertainty (N-L) further improves the uncertainty model.
    \label{tab:csd_log_likelihoods}
    }   
\end{figure}

Table~\ref{tab:csd_log_likelihoods} shows the log likelihood on the test set (Eq.~(\ref{eq:ll})) for different sub-sample sizes, before scaling with the maximum likelihood estimation scheme and after scaling, using either a validation set or internally with the training set as described earlier. It also shows the GPR log likelihoods before and after scaling with the validation set. Note that scaling the GPR variances using the training set for internal validation is impossible, hence the corresponding cells are empty.

We remark here that $LL$ (Raw) likelihoods in the first column of Table~\ref{tab:csd_log_likelihoods} can be directly compared to each other, when our goal is to evaluate relative efficiency of different sub-sample sizes. This is due to the fact that such comparison is equivalent to likelihood-based model selection with an Akaike information criterion-type (AIC) penalty \cite{Akaike_1974_AIC}. Indeed, AIC penalties depend only on the problem's dimensionality and not on the sub-sample size, so comparing penalized likelihoods in the AIC framework coincides with comparing $LL$ likelihoods in the first column of the Table. 

This is a convenient feature of our approach, as, in general, values of log-likelihoods for different sub-problems cannot be directly compared to each other \cite{Birge_Gaussian_2001,Langovoy_2007}. Strictly speaking, the powerful machinery of maximum likelihood only guarantees good properties of the best solution, but does not always directly induce a quality scale to rank other solutions via the use of the likelihood function. 

Before scaling, the log likelihood shows considerable variability between resampling estimators obtained with different sample size.
It appears that the estimator based on sub-samples of 5k environments (i.e. one quarter of the training set) strikes the best balance between resampling from the correct distribution but with the wrong size (small sub-sample sizes), and resampling from the wrong distribution but with the right size (large sub-sample sizes).
Rescaling the uncertainty estimator leads to a substantially more stable, standardized version of the log-likelihood, and reduces greatly the impact of the sample size for RS estimators. Additionally, we observed that after rescaling, the GPR-PP estimator leads to noticeably higher log likelihood.

\begin{figure}[thbp]
    \centering
    \includegraphics[scale=1.00]{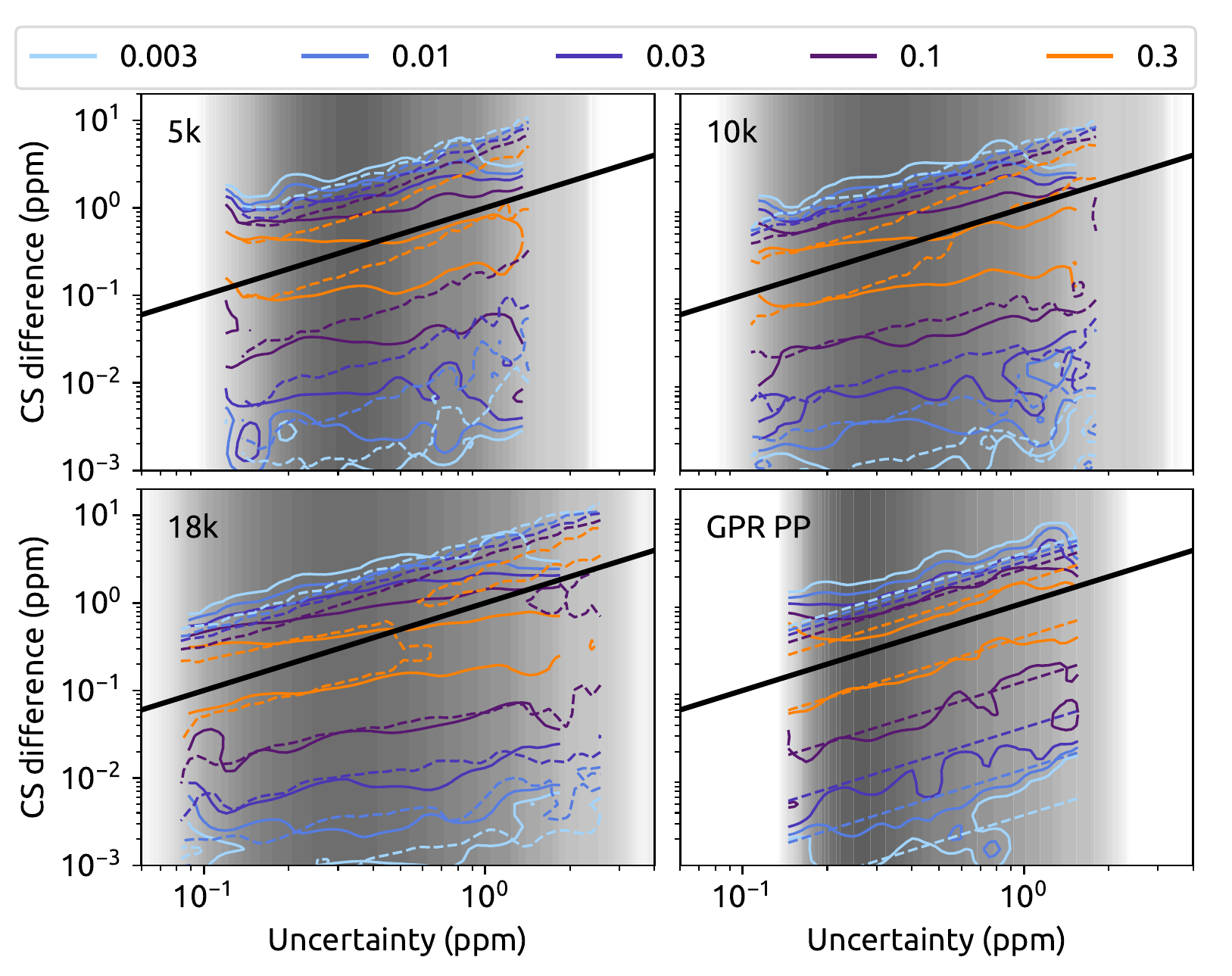}
    \caption{Distribution of \textsuperscript{1}H chemical shielding predictions. The solid line shows the distribution of $P\left(\ln \epsilon_t\middle|\ln \sigma\right)$, while the dashed line shows the distribution of $P\left(\ln \epsilon_m\middle|\ln \sigma\right)$ (see Eq.~\ref{eq:conditional}).
    The grayscale density plot corresponds to the marginal distribution of the predicted uncertainty $P(\ln \sigma)$.
    }
    \label{fig:csd_linear_contour}
\end{figure}

The log likelihood provides a measure of the accuracy of the uncertainty estimation that is quantitative but hardly intuitive. 
To provide a more straightforward representation of the accuracy of an uncertainty estimator, we observe that in an ideal scenario, the distribution of actual errors relative to the reference should match the distribution of the predictions of RS models around their mean. The equality of the distributions should be true for an arbitrarily-selected subset of the test set. 
Based on this observation, we computed the distribution of actual errors $\epsilon_t(\mathcal{X}_n)=\left|y(\mathcal{X}_n)-y_n \right|$ conditioned on the value of the predicted uncertainty $\sigma(\mathcal{X}_n)$, and the distribution of model errors $\epsilon_m(\mathcal{X}_n)=\left|y^{(i)}(\mathcal{X}_n)-y(\mathcal{X}_n) \right|$. 
Given that the predicted (and actual) errors can span a broad range, we computed the conditional on a log scale, e.g.
\begin{equation}
P\left(\ln\epsilon\middle|\ln\sigma\right)=
P\left(\ln\epsilon,\ln \sigma\right)/P\left(\ln\sigma\right).
\label{eq:conditional}
\end{equation}
The plots comparing the predicted and actual error distributions from the re-scaled estimators are shown in Fig.~\ref{fig:csd_linear_contour}.
One sees in all cases there is a good qualitative agreement between the distribution of the model (which is Gaussian by construction for the GPR model, and in some cases strongly non-Gaussian for RS estimators), with essentially none of the samples with large true errors being associated with a small $\sigma(\CX)$. 

\begin{figure}[thbp]
    \centering
    \includegraphics[scale=1.00]{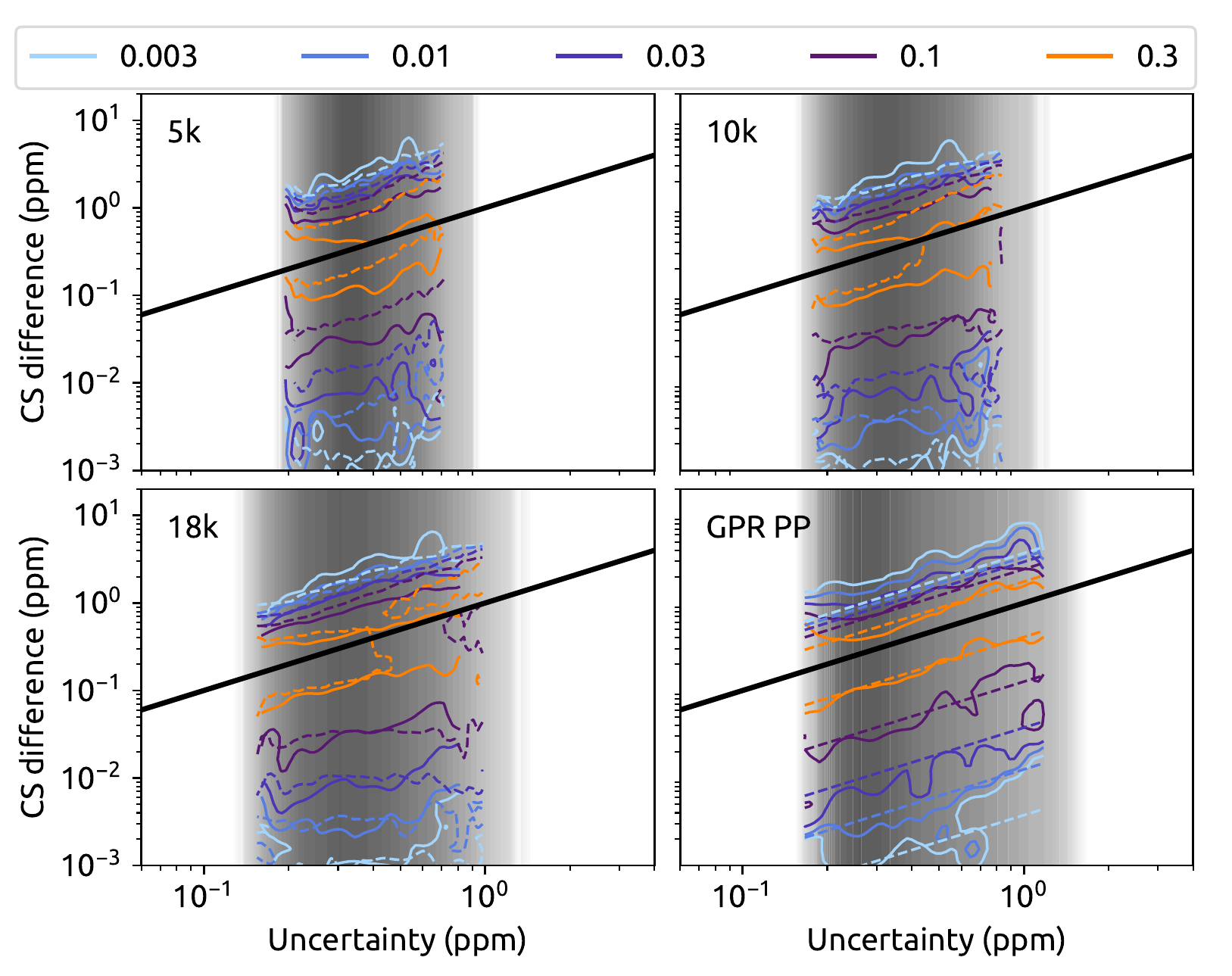}
    \caption{Distribution of \textsuperscript{1}H chemical shielding predictions. The solid line shows the distribution of $P\left(\ln\epsilon_t\middle|\ln\sigma\right)$, while the dashed line shows the distribution of $P\left(\ln\epsilon_m\middle|\ln\sigma\right)$ (see Eq.~\ref{eq:conditional}), including a non-linear scaling of the uncertainty corresponding to Eq.~\eqref{eq:rs-nonlinear-scaling}.
    The grayscale density plot corresponds to the marginal distribution of the predicted uncertainty $P(\ln\sigma)$.}
    \label{fig:csd_contour}
\end{figure}

On the other hand, there are also substantial differences between the various estimators. An obvious difference is the range spanned by the predicted $\sigma(\CX)$. One could argue that - for a given LL - the model that spans the broader range of values is the most useful, as it yields better resolution between more or less trustworthy predictions. 
From this point of view, large-sample-size RS estimators appear to be superior, spanning almost two orders of magnitude in the value of $\sigma(\CX)$. 
The GPR PP model, however, clearly displays the best agreement between predicted and actual error distributions, which is consistent with the higher LL. 

Looking more carefully at the distributions for the RS models, one can see that the actual errors tend to increase monotonically as a function of $\sigma(\CX)$, even though they do not follow the trend predicted by the sample distribution. This suggests that the performance of the estimator can be improved by introducing a more general transformation of $\sigma(\CX)$, and optimizing the parameters with a cross-validation procedure. 
As an example of the application of this idea, we define
\begin{equation}
{y}^{(i)}(\CX_n) \rightarrow y(\CX_n) + \alpha \left[{y}^{(i)}(\CX_n) -y(\CX_n)\right]\sigma(\CX_n)^{\gamma-1}.
\label{eq:rs-nonlinear-scaling}
\end{equation}
This transformation does not change the mean of the models, but generates a  non-linearly transformed $\sigma(\CX_n)\rightarrow\alpha\sigma(\CX_n)^{\gamma}$. 
The scaling $\alpha$ and exponent $\gamma$ are optimized by cross-validation. As shown in Fig.~\ref{fig:csd_contour} this procedure improves the agreement between the RS and the actual error distribution, and brings the LL close to the level of the GPR PP estimator (see also Table~\ref{tab:csd_log_likelihoods}). The observed improvement in fit hints at the possibility that the data are a better match to a light heavy-tailed process \cite{Gnedenko_Kolmogorov_1968} rather than to a standard Gaussian process. This is entirely plausible due to the high complexity of molecular models and high dimensionality of related data.  

\subsection{Prediction of QM9 formation energies}

\begin{figure}[H]
    \centering
    \begin{tabular}{l r r r r r } 
    \hline\hline
    Size &  \multicolumn{1}{c}{LL (Raw)}   & \multicolumn{1}{c}{LL (CV)} & \multicolumn{1}{c}{LL (N-L)} & \multicolumn{1}{c}{LL (Int.)} & \multicolumn{1}{c}{LL (Int., N-L)}\\
    \hline
    1k  &    4.310  & 4.651  & 4.647 & 4.654  & 4.652 \\
    5k  &    4.655  & 4.661  & 4.673 & 4.658  & 4.673 \\
    10k &    4.319  & 4.670  & 4.680 & 4.634  & 4.665 \\
    15k &    2.928  & 4.684  & 4.690 & 4.553  & 4.631 \\
    18k &    -0.807 & 4.681  & 4.686 & 4.462  & 4.568 \\
    GPR PP & 4.349   & 4.374 & 4.421 & N/A    & N/A    \\  
    \hline
    \end{tabular}
    \captionof{table}{Log likelihood (LL) of predictions on the test set for different sub-sample sizes. After scaling the variances through maximum likelihood estimation (internally or on the validation set), the final log likelihood is insensitive to the sub-sample size. A non-linear scaling of the uncertainty further improves the uncertainty model.
    \label{tab:qm9_log_likelihoods}
    }   
\end{figure}

Table~\ref{tab:qm9_log_likelihoods} is the analogue of Table~\ref{tab:csd_log_likelihoods} for the QM9 data set. The same trend is observed as for the CSD data set with the 5k sub-sampling estimator as the most reliable, before scaling through maximum likelihood estimation.
Despite the differences between predicting chemical shieldings (a local property) and formation energies (a global property), we see again that the (non-linear) scaling procedure with a validation set makes the GPR and the sub-sample uncertainty estimators more or less equally reliable.
The GPR estimator is consistently worse than the sub-sample estimators, however, even after scaling. This underscores the importance of having an objective way of comparing different uncertainty models, so that -- barring considerations related to the computational cost -- the best estimator can be selected by cross-validation.

\begin{figure}[t]
    \centering
    \includegraphics[scale=1.00]{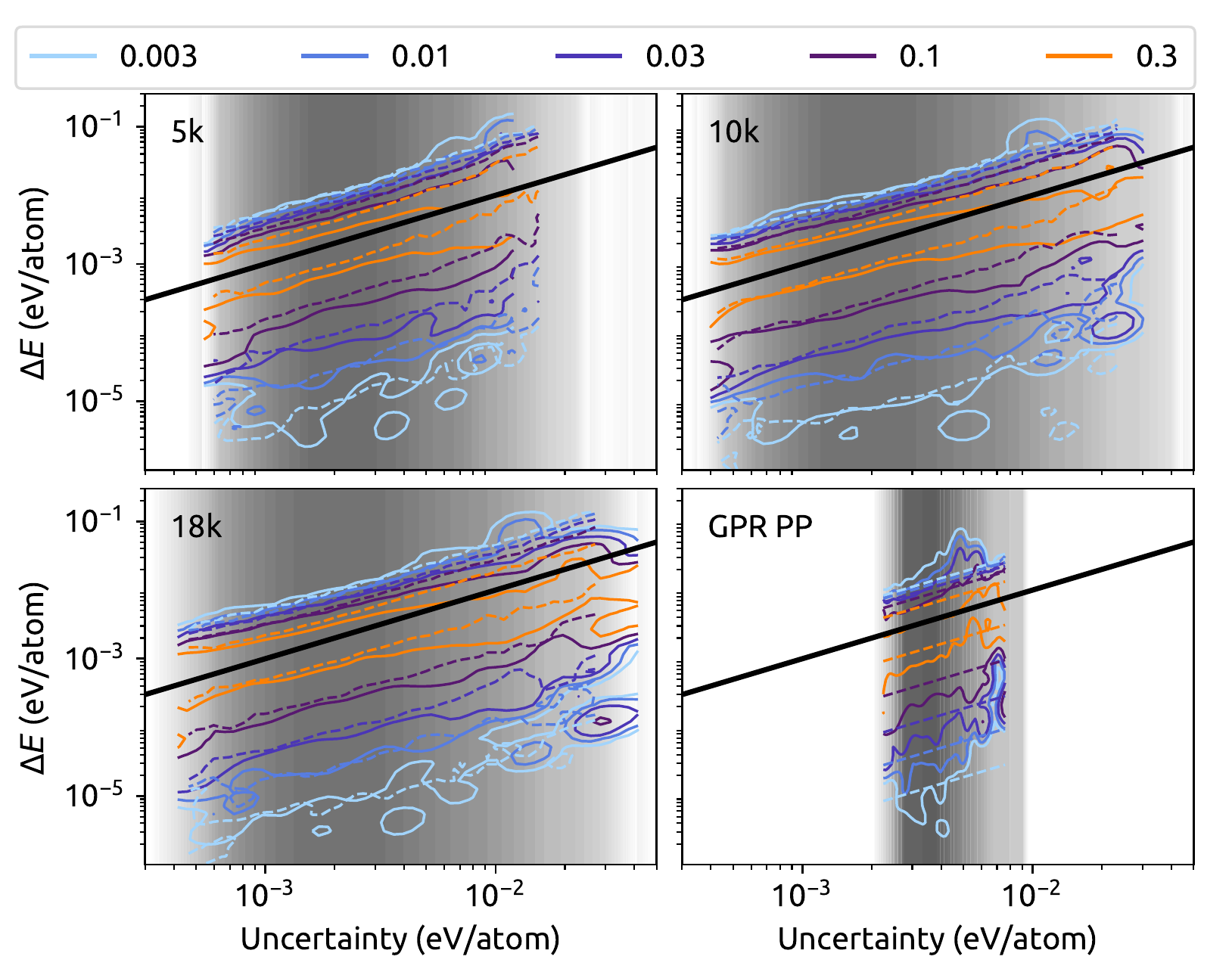}
    \caption{Distribution of formation energy differences. The solid line shows the distribution of $P\left(\ln \epsilon_t\middle|\ln \sigma\right)$, while the dashed line shows the distribution of $P\left(\ln \epsilon_m\middle|\ln \sigma\right)$ (see Eq.~\ref{eq:conditional}).
    The grayscale density plot corresponds to the marginal distribution of the predicted uncertainty $P(\ln\sigma)$.}
    \label{fig:qm9_linear_contour}
\end{figure}

Figure~\ref{fig:qm9_linear_contour} is the QM9 analogue of Figure~\ref{fig:csd_linear_contour}, reporting a more analytical representation of the correspondence between predicted and actual errors for the non-linearly scaled models. Again, the fundamental assumption of sub-sampling -- that the sub-samples are to the reference what the reference is to the target -- appears to be reliable. As for the CSD chemical shielding results, we found the quality of this agreement to be roughly the same, regardless of the sub-sample size. In this case, even after non-linear scaling, the GPR estimator yields a very narrow uncertainty distribution, while the RS models accurately predict the uncertainty over a span of two orders of magnitude.

\section{Conclusions}

We have presented a scheme to obtain an inexpensive and reliable estimate of the uncertainty associated with the predictions of a machine-learning model of atomic and molecular properties. The scheme is based on sub-sampling and sparse Gaussian Process Regression.
We have investigated the reliability of this approach for two applications: the prediction of \textsuperscript{1}H NMR chemical shieldings in organic crystals and the prediction of formation energies of small organic molecules. In both cases we found the sub-sampling estimator to be reliable on the basis of log likelihood results and the good agreement between the true and predicted distribution of errors on a test set. Besides the improvement of the method in comparison to the standard GPR uncertainty estimate from the point of view of computational scaling, it appears that the RS models remain accurate even when they are heavily compressed through the PP approach, which is encouraging since model compression is often required to accelerate model training and inference.

The framework we introduce to optimize the uncertainty model based on an objective measure of the accuracy of uncertainty estimation can be also applied easily to other machine learning schemes, e.g. neural networks etc., even though for many of these schemes the evaluation of the uncertainty may come at a substantial cost, contrary to the case of a sparse GPR model. 
Besides the computational savings, the fact that the sub-sampling models generate an ensemble of predictions makes it trivial to predict uncertainties in derived properties, that are obtained by linear or non-linear combination of multiple predictions. 

An accurate and inexpensive uncertainty estimation opens up promising research directions in terms of training-set optimization, and the development of active-learning strategies~\cite{li+15prl}. It also suggests the possibility of developing committee models~\cite{Breiman1996,Tresp2000} in which multiple ML predictions are performed, differing in the definition of the kernel and/or regularization. These can then be combined based on their uncertainty estimation, which would allow for instance to develop schemes that degrade gently to less accurate but more robust models when entering an extrapolative regime. 

\section{Acknowledgements}

MC and MJW were supported by the European Research Council under the European Union's Horizon 2020 research and innovation programme (grant agreement no. 677013-HBMAP).  FM and ML were supported by NCCR MARVEL, funded by the Swiss National Science Foundation.
Insightful discussion with Martin Jaggi, Federico Paruzzo and Albert Hofstatter are gratefully acknowledged. 

\providecommand{\latin}[1]{#1}
\makeatletter
\providecommand{\doi}
  {\begingroup\let\do\@makeother\dospecials
  \catcode`\{=1 \catcode`\}=2 \doi@aux}
\providecommand{\doi@aux}[1]{\endgroup\texttt{#1}}
\makeatother
\providecommand*\mcitethebibliography{\thebibliography}
\csname @ifundefined\endcsname{endmcitethebibliography}
  {\let\endmcitethebibliography\endthebibliography}{}

\end{document}